\documentclass{article}
\usepackage{ijcai19}

\usepackage{times}
\usepackage{soul}
\usepackage{url}
\usepackage[hidelinks]{hyperref}
\usepackage[utf8]{inputenc}
\usepackage[small]{caption}
\usepackage{graphicx}
\usepackage{amsmath}
\usepackage{amsthm}
\usepackage{booktabs}
\usepackage{algorithm}
\usepackage{algorithmic}
\usepackage{subcaption}
\usepackage[bottom]{footmisc}
\title{Dilated Convolution with Dilated GRU for Music Source Separation}

\author{
Jen-Yu Liu \and %\And
Yi-Hsuan Yang
\affiliations
Research Center for IT Innovation, Academia Sinica
\emails
jenyuliu.tw@gmail.com, yang@citi.sinica.edu.tw
}

\begin{document}

\maketitle

\begin{abstract}
  Stacked dilated convolutions used in Wavenet have been shown effective for generating high-quality audios. By replacing pooling/striding with dilation in convolution layers, they can preserve high-resolution information and still reach distant locations. Producing high-resolution predictions is also crucial in music source separation, whose goal is to separate different sound sources while maintaining the quality of the separated sounds. Therefore, this paper investigates using stacked dilated convolutions as the backbone for music source separation. However, while stacked dilated convolutions can reach wider context than standard convolutions, their effective receptive fields are still fixed and may not be wide enough for complex music audio signals. To reach
  %even further
  information at remote locations, we propose to combine dilated convolution with a modified version of gated recurrent units (GRU) called the `Dilated GRU' to form a block. A Dilated GRU unit receives information from $k$ steps before instead of the previous step for a fixed $k$. This modification allows a GRU unit to reach a location with fewer recurrent steps and run faster because it can execute partially in parallel. We show that the proposed model with a stack of such blocks performs equally well or better than the state-of-the-art models for separating vocals and accompaniments.

\end{abstract}

\section{Introduction}
%https://www.overleaf.com/learn/latex/Questions/What_does_%22%5Cpdfendlink_ended_up_in_different_nesting_level_than_%5Cpdfstartlink%22_mean%3F

Music source separation has received much attention and has obtained impressive progress in recent years \cite{Stoter2018}. It has different use cases such as generating accompaniment from pop songs for karaoke \cite{Rafii2018}, separating specific sources as a pre-processing tool for other tasks such as music transcription \cite{Paulus2005}, and DJ-related applications \cite{VandeVeire2018}.

In recent years, neural network-based methods have obtained promising result for music source separation \cite{Nugraha2016,Uhlich2017,Takahashi2018,Liutkus2017,Stoter2018}. Among them, \cite{Nugraha2016} proposed a DNN architecture with fully-connected layers, and it is one of the first models based on neural networks for music source separation. Music source separation requires the model to produce high-resolution predictions. This has mostly been  achieved by either using encoder-decoder architecture with skip connections \cite{Jansson2017,Takahashi2017,Takahashi2018,Liu2018,StollerED18}, or recurrent neural networks \cite{Uhlich2017}.

We notice that the stacked dilated convolutions used in Wavenet \cite{Oord2016} might also work well for this purpose. In Wavenet, the kernels in convolution layers are dilated more and more as the network goes deeper, so the entire network can access neighboring information as well as distant information, depending on how many dilated convolution layers are stacked. 

Dilated convolutions are used in audio generation \cite{Oord2016}, machine translation \cite{Kalchbrenner2017}, speech recognition \cite{Sercu2016}, semantic segmentation \cite{Yu2015}, and video generation \cite{Kalchbrenner2017}. To the best of our knowledge, dilated convolutions have not been used for audio regression problems such as music source separation.

Music audio signals are usually very long, even using spectrogram as the feature. For example, a 3-minute audio under 44,100 Hz sampling rate and 1,024-sample hop size for short-time Fourier transform has 7,752 frames in its spectrogram representation. It is not easy to stack enough dilated convolutions to reach that far with limited computational resources.

We propose to combine a recurrent layer, specifically a gated recurrent unit (GRU) \cite{Cho2014,Chung2014}, and a dilated convolution to form a block. A GRU can in theory reach very far away in one layer if the information does not decay too fast. As aforementioned, music audio signals can be very long. This can sometimes be a problem for a recurrent layer like GRU, because it has to process its input sequence sequentially. 

We use the \emph{Dilated GRU} to alleviate this problem. A Dilated GRU unit receives information from $k$-step before instead of the previous step for a fixed $k$. This modification allows a GRU unit to reach a location with fewer recurrent steps and run faster because it can execute partially in parallel. The dilated version of recurrent layers is also used in other contexts. For example, \cite{Chang2017} stacked multiple recurrent layers with increasing dilations for speaker identification, while \cite{Vezhnevets2017} used them as managers in reinforcement learning, both of which are pure RNN architectures. In contrast, we combine them with dilated grouped convolutions to form processing blocks. We call these blocks the  `\underline{D}ilated recurrent-\underline{D}ilated convolution' (D2) blocks.

We conduct extensive experiments to verify the capability of the proposed model, as well as the relative importance of its components. 
We also investigate how the D2 blocks work. % in the proposed model.
Our evaluation shows that our model (GRU dilation=1) outperforms the state-of-the-art model  \cite{Takahashi2018} by 0.25 dB and 0.57 dB in the signal-to-distortion ratio (SDR) for vocal separation and accompaniment, respectively.

\section{Proposed model}
Our model works on the magnitude of spectrograms. For a $D\times T$ tensor $M_{mix}$ of the mixture magnitude, the goal of source separation is to predict a $D\times T$ tensor $M_s$ for each source $s\in S$, where $D$, $T$, and $S$ denotes the feature dimension, the sequence length, and the set of sources, respectively. Our model takes $M_{mix}$ as the input, and predicts all the source tensors $M_s$ at once.

\subsection{Dilated GRU}

GRU\footnote{We use the GRU version implemented by PyTorch (\url{https://pytorch.org/docs/stable/nn.html\#gru}), which is slightly different from the original version.} \cite{Cho2014,Chung2014} is a popular design choice of recurrent layers. In this paper, we use a modified version of GRU where the unit at a temporal point receives information from $k$-step before instead of the previous step for a fixed $k$. The same idea of dilating the temporal connections of an RNN has also been used by \cite{Vezhnevets2017,Chang2017}. 

\begin{figure}
	\centering
	\includegraphics[width=\columnwidth]{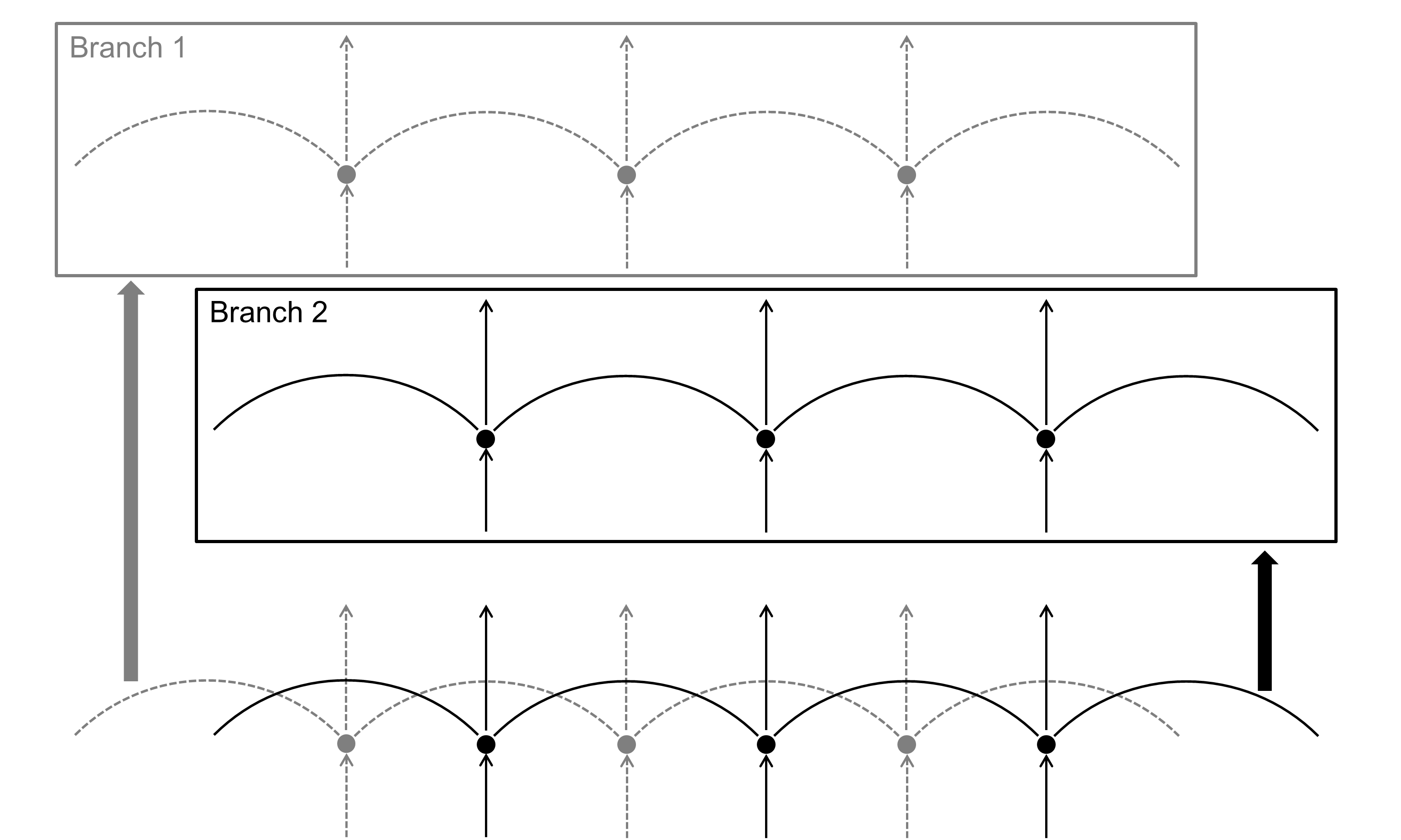}
	\caption{Parallel branches in Dilated GRU. The locations with odd indices and even indices can be processed in parallel with dilation 2.}
	\label{fig:parallel}
\end{figure}

Dilated GRU with dilation $k$ involves the following operations:
\begin{align*}
    r_t &= \sigma(W_{ir} x_t + b_{ir} + W_{hr} h_{(t-k)} + b_{hr})\,, \\
    z_t &= \sigma(W_{iz} x_t + b_{iz} + W_{hz} h_{(t-k)} + b_{hz})\,, \\
    n_t &= \tanh(W_{in} x_t + b_{in} + r_t (W_{hn} h_{(t-k)}+ b_{hn}))\,, \\
    h_t &= (1 - z_t) n_t + z_t h_{(t-k)}\,,
\end{align*}
where $W$ stands for matrices, $b$ the bias terms, and $\sigma$ the sigmoid function. $r$, $z$, $n$ and $h$ are all vectors. The temporal indices are partitioned into $k$ disjoint sets, meaning that they can be processed independently in parallel.  Figure \ref{fig:parallel} provides an illustration.

\subsection{Dilated Recurrent-Dilated Convolution Block (D2 block)}\label{sec:d2block}

In contrast to the a conventional convolution, the kernel of a dilated convolution is dilated so that it takes as input the further neighbors instead of the immediate neighbors. 

\begin{figure}
	\centering
	\begin{subfigure}{\columnwidth}
		\centering
		\includegraphics[width=\columnwidth]{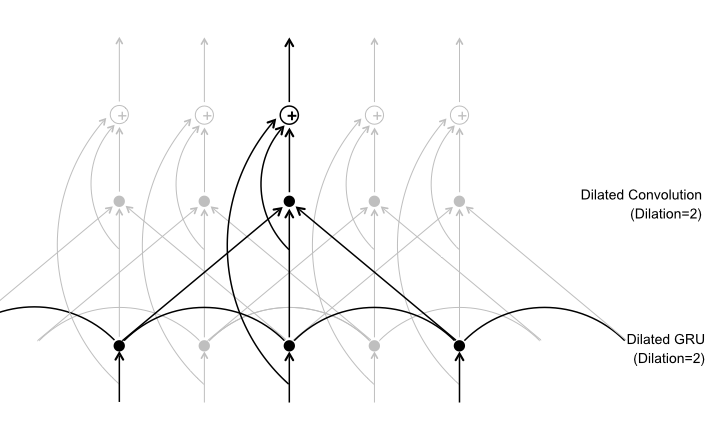}
		\caption{Temporal view}
		\label{fig:block_temporal}
	\end{subfigure}
	\begin{subfigure}{\columnwidth}
		\centering
		\includegraphics[width=\columnwidth]{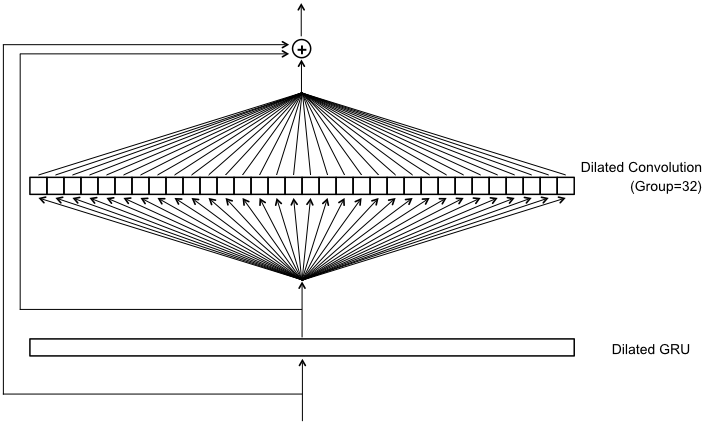}
		\caption{Channel view}
		\label{fig:block_channel}
	\end{subfigure}
	\caption{Dilated GRU-Dilated convolution block (D2 block). A Dilated GRU layer is followed by a dilated convolution with groups. The input of the block, the output of Dilated GRU and the output of the dilated convolution are added together to the output of the block. Two views of the proposed architecture are shown, since the architecture contains temporal connections (better seen in the Temporal view) as well as grouped channels in convolutions (better seen in the Channel view).}
	\label{fig:block_design}
\end{figure}

We propose to stack a dilated grouped convolution on top of a Dilated GRU layer with bidirectional connections. The input to the block, the output of the Dilated GRU and the output of the dilated grouped convolution are summed together to form the output of the block. This is illustrated  in Figure \ref{fig:block_design}. Leaky ReLU and weight normalization \cite{Salimans2016} are applied to all the convolution layers.

In stacked dilated convolutions, the model could potentially pay less attention to neighboring locations as the dilation in convolution layers increases. We hope to alleviate this problem by using vertical skip connections from the block input to the block output, as also shown in Figure \ref{fig:block_design}.

The grouping in a convolution layer is also used in architectures such as MobileNet \cite{Howard2017} and ShuffleNet \cite{Zhang2018}. It reduces the connections between input and output and hence reduces the memory usage and computation while still maintaining the capacity of the network. In one layer, the different groups have their own inputs and do not communicate with each other. Different groups cannot directly  communicate with each other, so the grouped convolutions are usually followed some mixing layers such as fully-connected layers \cite{Howard2017} or shuffle layers \cite{Zhang2018}. We do not use these extra layers in the proposed model, but instead delegate this communication task to the GRU layer, since the computation in GRU is fully-connected across channels. Therefore, the Dilated GRU at the beginning of each block will mix the information from different groups in addition to its other task of aggregating information through time.

\begin{figure}
	\centering
	\begin{subfigure}{\columnwidth}
		\centering
		\includegraphics[width=\columnwidth]{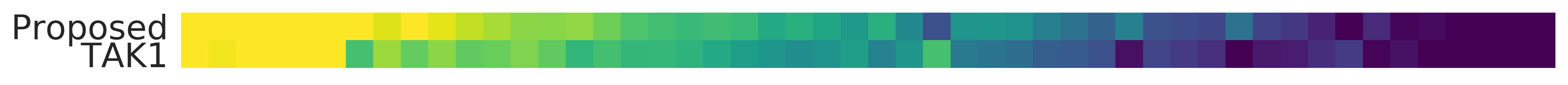}
		\caption{Vocals}
		\label{fig:heatmap_vocals}
	\end{subfigure}
	\begin{subfigure}{\columnwidth}
		\centering
		\includegraphics[width=\columnwidth]{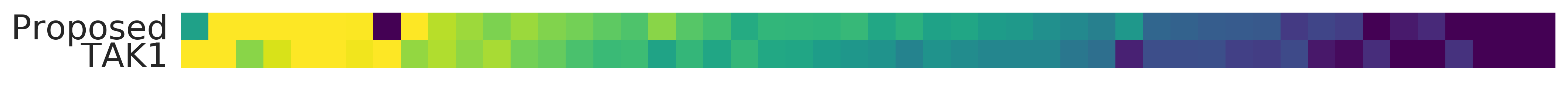}
		\caption{Accompaniment}
		\label{fig:heatmap_acc}
	\end{subfigure}
	\caption{Track-by-track comparison of the 50 test songs. The SDR of a track is represented by a block colored with a color on the yellow-green-black color map (high to low). The `Proposed' represents the model of GRU dilation 1 trained with 20-sec clips.}
	\label{fig:heatmap}
\end{figure}

\subsection{Full model}

\begin{figure}
	\centering
	\includegraphics[width=0.8\columnwidth]{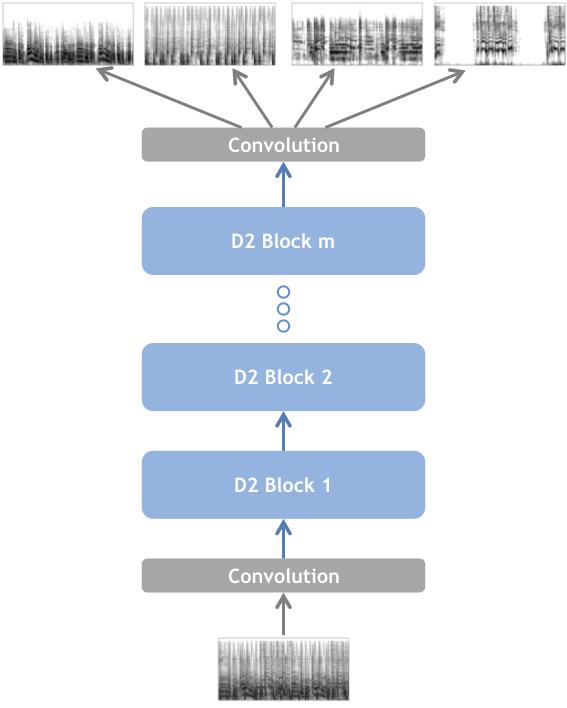}
	\caption{Network architecture of the proposed model. A convolution layer processes the input feature map, followed by a stack of D2 blocks. A convolution layer processes the output of the last D2 block and outputs all the predicted separations.}
	\label{fig:model}
\end{figure}

The full model is shown in Figure \ref{fig:model}. The input is first processed by a convolution layer. Then, several D2 blocks are stacked together. By the end, the output of the last D2 block is processed by a convolution. It then outputs the separation predictions of all the sources at once.

Most top models in SiSEC2018 use certain forms of denoising auto-encoders that down-sample and up-sample along the temporal axis in the encoders and decoders respectively \cite{Jansson2017,Takahashi2017,Takahashi2018,Liu2018}. They also use symmetric skip connections connecting a pair of encoder and decoder to compensate the loss of high-resolution information due to down-sampling. In contrast, our models maintain the temporal resolution without down-sampling and up-sampling during the whole process, so the high-resolution information will not be lost. The blocks can be easily stacked without considering the symmetry of skip connections across different blocks in this approach.

More details of our model can be found in Table \ref{tab:model}.

\begin{table}
\centering
\begin{tabular}{llll}
\toprule
Unit                & Spec.     & Input & Output \\
\midrule
Input convolution   & 1D Conv (3, 1)& 2049  & 2048\\
\midrule
Block 1             & Dilated GRU      & 2048  & 2048\\
                    & 1D Conv (3, 2)& 2048  & 2048\\
\midrule
Block 2             & Dilated GRU      & 2048  & 2048\\
                    & 1D Conv (3, 4)& 2048  & 2048\\
\midrule
Block 3             & Dilated GRU      & 2048  & 2048\\
                    & 1D Conv (3, 8)& 2048  & 2048\\
\midrule
Output convolution  & 1D Conv (3, 1)& 2048  & 4$\times$2049\\
\bottomrule
\end{tabular}
\caption{Model specification. The two numbers in the parentheses after `1D Conv' are kernel size and dilation size, respectively. All the 1D convolution layers in the Blocks 1, 2 and 3 have 32 groups, where each group has 64 input channels and 64 output channels.}
\label{tab:model}
\end{table}

\section{Evaluation}\label{sec:eval}

\subsection{Evaluation Setup}\label{sec:setting}
We evaluate the proposed model for music source separation on the MUSDB18 dataset\footnote{\url{https://sigsep.github.io/datasets/musdb.html}} used in SiSEC2018 \cite{Stoter2018}. MUSDB18 contains 100 songs for training and 50 songs for evaluation, all with 44,100 Hz sampling rate. Each song contains the source audios of `vocals,' `drums,' `bass' and `other.' The evaluation is conducted by using the evaluation package provided by the SiSEC2018 organizers.\footnote{\url{https://github.com/sigsep/sigsep-mus-eval}} The evaluation metrics are computed by taking the median over all the tracks in the evaluation set as done in the official report \cite{Stoter2018}, using the SiSEC analysis package.\footnote{\url{https://github.com/sigsep/sigsep-mus-2018-analysis/blob/master/sisec-2018-paper-figures/boxplot.py}} %Signal-to-distortion ratio (SDR) is reported in most tables.

We evaluate the performance of the models mainly with objective metrics commonly used in music source separation. Unless otherwise specified, we report the SDR in the tables. 

Our models use the log-scale magnitudes of the complex spectrogram as the feature. First, complex spectrograms are derived by applying a short-term Fourier transform (STFT) to waveforms, with 4,096-sample window size and 3/4 overlapping.
Then, the magnitudes of the complex spectrograms are computed. $log(1+\text{magnitude})$ of the mixture audios and the source audios are used as the inputs and the training targets, respectively. Mean square error is used as the loss function for updating the network, and Adam \cite{Kingma2014} is used to update the weights. As data augmentation has been found useful in the literature \cite{Takahashi2018,Uhlich2017,Liu2018}, we use data augmentation in the training process by randomly shuffling the audio clips in each source and then collecting the audio clips from the four sources in the shuffled orders. A mixture clip is formed by summing the collected source clips. The source audio clips are shuffled at the beginning of every epoch. Batch sizes 20 and 5 are used for 5-sec and 20-sec training respectively so that they have roughly the same number of weight updating in training. 1/10 of the training set is used for validation. The weights in the epoch with the best validation loss during 500-epoch training are kept for a model.

To convert the outputs of the model to waveforms, the above process is reversed. The phases of the mixture complex spectrograms are used with the predicted spectrogram magnitudes to construct the complex spectrogram. Before converting back to waveforms, multi-channel Wiener filter is applied to the complex spectrograms as widely done in recent source separation systems \cite{Nugraha2016,Uhlich2017,Takahashi2018,Liu2018}.

\subsection{Comparison with Participants of SiSEC2018}\label{sec:sisec}

\begin{table*}
\centering
\begin{tabular}{lllllll}
\toprule
Model   & Description   & vocals& drums	& bass	& other	& accomp. \\
\midrule
TAK1 \cite{Takahashi2018}   &   & 6.60  & \textbf{6.43}  & \textbf{5.16}  & 4.15  & 12.83\\
UHL2 \cite{Uhlich2017}      &   & 5.93  & 5.92  & 5.03  & 4.19  & 12.23\\
JY3 \cite{Liu2018}          &   & 5.74  & 4.66  & 3.67  & 3.40  & 12.08\\
MDL1 \cite{Mimilakis2018}   &   & 4.02  & NA    & NA    & NA    & 9.92\\
RGT1 \cite{Roma2018}        &   & 3.85  & 3.44  & 2.70  & 2.63  & NA\\
\midrule
Stacked dilated convolutions    & 20-sec training    & 5.34 $\pm$ 0.17 & 5.20 $\pm$ 0.05       & 3.72 $\pm$ 0.10       & 3.33 $\pm$ 0.12       & 11.88 $\pm$ 0.05\\
Stacked dilated convolutions    & 5-sec training     & 5.56 $\pm$ 0.07 & 5.35 $\pm$ 0.19       & 3.76 $\pm$ 0.04       & 3.52 $\pm$ 0.07       & 11.95 $\pm$ 0.09 \\
\midrule
Proposed, GRU dilation 2& 20-sec training& 6.76 $\pm$ 0.06  & 5.85 $\pm$ 0.07 & 4.84 $\pm$ 0.09 & 4.49 $\pm$ 0.08 & 13.19 $\pm$ 0.02 \\
Proposed, GRU dilation 2&  5-sec training& 6.78 $\pm$ 0.05  & 5.66 $\pm$ 0.09 & 4.96 $\pm$ 0.03 & 4.48 $\pm$ 0.09 & 13.22 $\pm$ 0.05 \\
Proposed, GRU dilation 1& 20-sec training& \textbf{6.85} $\pm$ 0.04  & 5.86 $\pm$ 0.12 & 4.86 $\pm$ 0.05 & \textbf{4.65} $\pm$ 0.04 & \textbf{13.40} $\pm$ 0.11\\
Proposed, GRU dilation 1& 5-sec training& 6.81 $\pm$ 0.15  & 5.72 $\pm$ 0.06 & 4.58 $\pm$ 0.10 & 4.48 $\pm$ 0.11 & 13.26 $\pm$ 0.08\\

\bottomrule
\end{tabular}
\caption{Comparison with models in SiSEC2018 (in SDR). The table shows the top models in SiSEC2018 in the upper part, the baseline stacked dilated convolutions in the middle part, and the proposed models in the lower part. All the proposed models contain three blocks.}
\label{tab:sisec2018}
\end{table*}

Table \ref{tab:sisec2018} shows the performance of the proposed models and the top-performing models of SiSEC2018 \cite{Rafii2018}.\footnote{The raw data of the evaluation metrics are available at [Online] \url{https://github.com/sigsep/sigsep-mus-2018}} We show the top 5 models (one top model from each group) that are trained without extra data in SiSEC2018. In this subsection, each setting is run three times, and the mean score and standard deviation of the three runs is reported.

First, we can see that the model using stacked dilated convolutions alone without recurrent layers and skip connections already have performance comparable with the top-3 model JY3 in SiSEC2018. This verifies our intuition that stacked dilated convolutions are strong not only in generation problems but also in music source separation.

The best performance in both `vocals' and `accompaniment' is achieved by the proposed model with dilation 1 trained with 20-sec clips. In general, models trained with longer and shorter clips have similar performance.

Dilation 2 and dilation 1 are both strong models. The performance difference between the proposed models with dilation 1 and dilation 2 are not as large as the difference between the difference between the proposed models and TAK1. In practice, the proposed model with dilation 2 has the benefit of parallel computation.

Our models are strong in `vocals,' `other,' and overall `accompaniment.' They perform slightly worse in `drums' and `bass' compared to TAK1 and UHL2. Our conjecture is that the different sources are competing for resources in the model because we train one model for all four sources. The sounds in `vocals' and `other' are usually louder than `drums' and `bass,' so they are weighted more in the loss function. In contrast, TAK1 and UHL2 do not have this problem because they train one model for each source \cite{Takahashi2018,Uhlich2017}.
%so the sources do not have to compete for resources. 
This can be a trade-off between resources and performance.

In Table \ref{tab:sota}, we compare our models with TAK1, including other performance metrics in addition to SDR. Our models consistently outperform TAK1 in `vocals' and `other' in all metrics. We also show the track-by-track comparison in Figure \ref{fig:heatmap}. In general, our proposed model performs better in vocals in almost all tracks, while our model and TAK1 have their own advantages in accompaniment in different tracks. 

\begin{table}
\centering
\setlength{\tabcolsep}{1pt}
\small
\begin{tabular}{llllll}
\toprule
\textbf{SDR}    & vocals& drums	& bass	& other	& accomp. \\
\midrule
TAK1 	    & 6.60          & \textbf{6.43} & \textbf{5.16} & 4.15          & 12.83  \\
D=2  & 6.76 $\pm$ .06  & 5.85 $\pm$ .07 & 4.84 $\pm$ .09 & 4.49 $\pm$ .08 & 13.19 $\pm$ .02\\
D=1  & \textbf{6.85} $\pm$ .04  & 5.86 $\pm$ .12 & 4.86 $\pm$ .05 & \textbf{4.65} $\pm$ .04 & \textbf{13.40} $\pm$ .11\\
\midrule
\midrule
\textbf{SIR}    & vocals& drums	& bass	& other	& accomp. \\
\midrule
TAK1 	    & 14.37         & 11.81         & \textbf{10.54}& 6.41          & 16.69  \\
D=2  & \textbf{14.60} $\pm$ .30 & 11.65 $\pm$ .13 & 9.44 $\pm$ .13 & 7.37 $\pm$ .27 & 18.22 $\pm$ .24\\
D=1  & 14.33 $\pm$ .31 & \textbf{12.26} $\pm$ .21 & 9.20 $\pm$ .40 & \textbf{7.58} $\pm$ .13 & \textbf{18.43} $\pm$ .31\\
\midrule
\midrule
\textbf{SAR}    & vocals& drums	& bass	& other	& accomp. \\
\midrule
TAK1 	    & 6.37          & \textbf{6.64} & 5.69          & 4.83          & \textbf{14.08}  \\
D=2  & 6.50 $\pm$ .18  & 6.05 $\pm$ .12 & \textbf{5.89} $\pm$ .19 & \textbf{4.98} $\pm$ .02 & 13.63 $\pm$ .12\\
D=1  & \textbf{6.56} $\pm$ .05  & 6.13 $\pm$ .23 & 5.82 $\pm$ .10 & 4.88 $\pm$ .06 & 13.72 $\pm$ .18\\
\midrule
\midrule
\textbf{ISR}    & vocals& drums	& bass	& other	& accomp. \\
\midrule
TAK1 	    & 11.56         & \textbf{12.02}& 9.92          & \textbf{9.86} & \textbf{22.56}  \\
D=2  & 13.24 $\pm$ .05 & 10.62 $\pm$ .24 & 9.63 $\pm$ .28 & 9.56 $\pm$ .17 & 22.38 $\pm$ .13\\
D=1  & \textbf{13.50} $\pm$ .10 & 10.69 $\pm$ .20 & \textbf{10.20} $\pm$ .31 & 9.58 $\pm$ .12 & 22.39 $\pm$ .31\\
\bottomrule

\end{tabular}
\setlength{\tabcolsep}{6pt}
%\caption{Comparison with the state-of-the-art (TAK1)}
\caption{Performance comparison with the state-of-the-art model (TAK1) \protect\cite{Takahashi2018} using SDR and other metrics. Our models (GRU dilations 1 and 2) are trained with 20-sec subclips. `D' represents dilation.}
\label{tab:sota}

\end{table}

\subsection{Ablation Study}

We consider ablated versions of our model in this evaluation.
%In this subsection, we ablate our model in different aspects. 
Specifically, we use the model with dilation 2 and trained with 5-sec subclips as the baseline. The score of one training run for each setting is reported in this subsection. We will refer to the model in Section \ref{sec:d2block} and Table \ref{tab:model} as the proposed model.

First, we compare different block designs. The first one is the proposed block, the second one, `Dense,' is similar to the proposed block with an additional skip connection adding the input of the block to the output of Dilated GRU, the third one, `Residual,' has a residual connection for each layer including Dilated GRU and convolutions, and the fourth one has the same skip connections as the proposed block but the dilated convolution layer is before the Dilated GRU. To investigate the effectiveness of GRU, we replace GRU with a convolution in the fifth row. The performance of the variants are presented in Table \ref{tab:block_variants}. The proposed one has the best performance.

\begin{table}
\centering
\setlength{\tabcolsep}{4pt}
\begin{tabular}{lrrrrr}
\toprule
            & vocals& drums	& bass	& other	& accomp. \\
\midrule
DGRU-DGConv & \textbf{6.74}  & 5.71  & \textbf{5.00}  & \textbf{4.61}  & \textbf{13.25} \\
*(Dense)    & 6.60  & 5.71  & 4.72  & 4.36  & 13.19\\
*(Residual) & 6.50  & \textbf{5.86}  & 4.83  & 4.26  & 13.16 \\
DGConv-DGRU & 6.59  & 5.64  & 4.42  & 4.17  & 13.15\\
Conv-DGConv & 5.81  & 5.28  & 3.97  & 3.65  & 12.24\\
\bottomrule
\end{tabular}
\caption{Performance comparison (in SDR) of different block designs, using dilation 2 and 3 blocks. `DGRU' represents Dilated GRU with dilation 2, `DGConv' represents the dilated grouped convolution described in Section \ref{sec:d2block}.`DGRU-Conv' is the one presented in Table \ref{tab:model}. `Conv-DGConv' replaces all GRUs with non-grouped 1D convolution with kernel size 1.}
\label{tab:block_variants}
\end{table}

Second, we compare different GRU dilations in Table \ref{tab:numjumps}. Dilation 1 and 2 perform better either in vocals or accompaniment than dilation 4. \cite{Chang2017} constructed a recurrent neural network with increasing dilations in the recurrent layers. We also evaluate our model with increasing dilations, but found no improvement as shown in the last row of Table \ref{tab:numjumps}.

To evaluate the running speed, we run the model with one batch 10 times for each setting and compute the average running time. The average are 423$\pm$4.90, 340$\pm$8.91, and 327$\pm$10.2 ms for models with dilation 1, dilation 2, and dilation 4, respectively. The running time is reduced about 20\% from dilation 1 to dilation 2, but there is only marginal reduction from dilation 2 to dilation 4.

\begin{table}
\label{tab:jump}
\centering
\begin{tabular}{lrrrrr}
\toprule
Dilation    & vocals& drums	& bass	& other	& accomp. \\
\midrule
1       & \textbf{6.99} & \textbf{5.79} & 4.55  & 4.49  & 13.22 \\
2       & 6.74          & 5.71          & \textbf{5.00}  & \textbf{4.61}  & \textbf{13.25} \\
4       & 6.72          & 5.48          & 4.41  & 4.22  & 13.07 \\
2, 4, 8 & 6.53          & 5.70          & 4.68  & 4.29  & 12.86\\
\bottomrule
\end{tabular}
\caption{SDR of the models with different GRU dilations, all with 3 blocks.}
\label{tab:numjumps}
\end{table}

Third, models with different number of D2 blocks are compared in Table \ref{tab:numblocks}. 3 blocks and 4 blocks have close results.

\begin{table}
\centering

\begin{tabular}{lrrrrr}
\toprule
\# blocks & vocals& drums	& bass	& other	& accomp. \\
\midrule
2   & 6.49          & 5.49          & 4.89          & 4.32  & 13.01 \\
3   & 6.74          & \textbf{5.71} & \textbf{5.00} & \textbf{4.61}  & \textbf{13.25} \\
4   & \textbf{6.81} & 5.64          & 4.84          & 4.50  & 13.26 \\
\bottomrule
\end{tabular}
\caption{SDR of variants of our model with different number of blocks, all with GRU dilation 2.}
\label{tab:numblocks}
\end{table}

\begin{figure*}
	\centering
	\begin{subfigure}{0.45\textwidth}
		\centering
		\includegraphics[width=\textwidth]{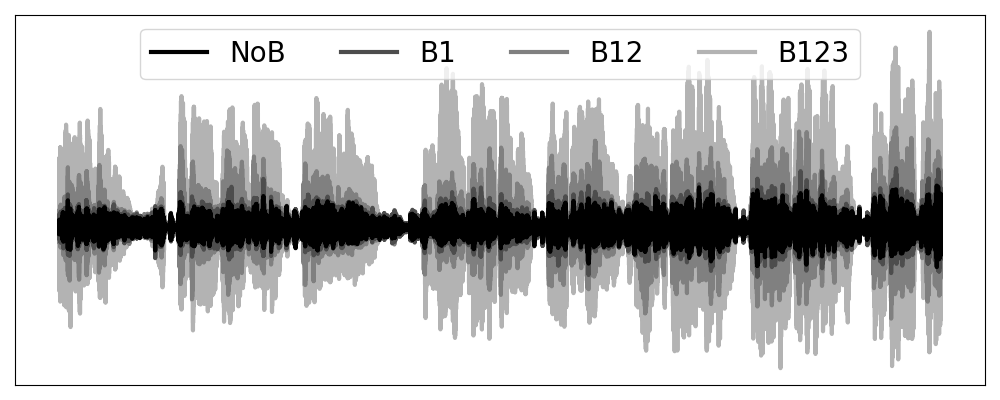}
		\caption{Vocals}
	\end{subfigure}%
	\begin{subfigure}{0.45\textwidth}
		\centering
		\includegraphics[width=\textwidth]{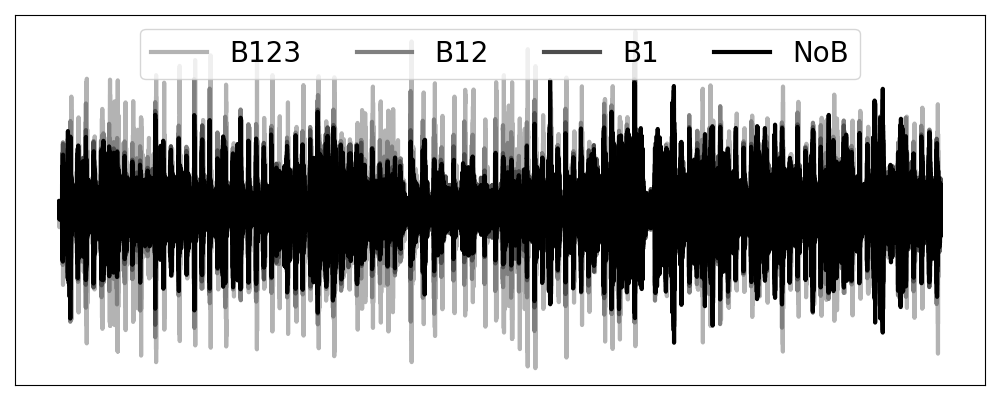}
		\caption{Drums}
	\end{subfigure}%
	\par\bigskip
	\begin{subfigure}{0.45\textwidth}
		\centering
		\includegraphics[width=\textwidth]{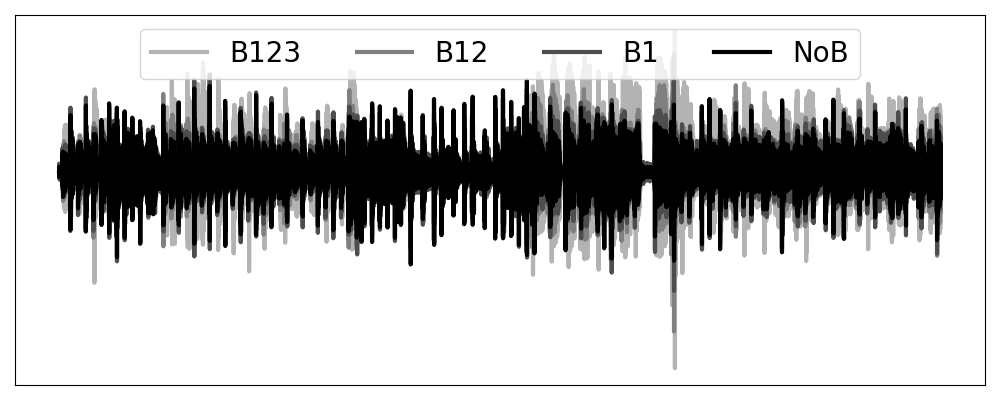}
		\caption{Bass}
	\end{subfigure}%
	\begin{subfigure}{0.45\textwidth}
		\centering
		\includegraphics[width=\textwidth]{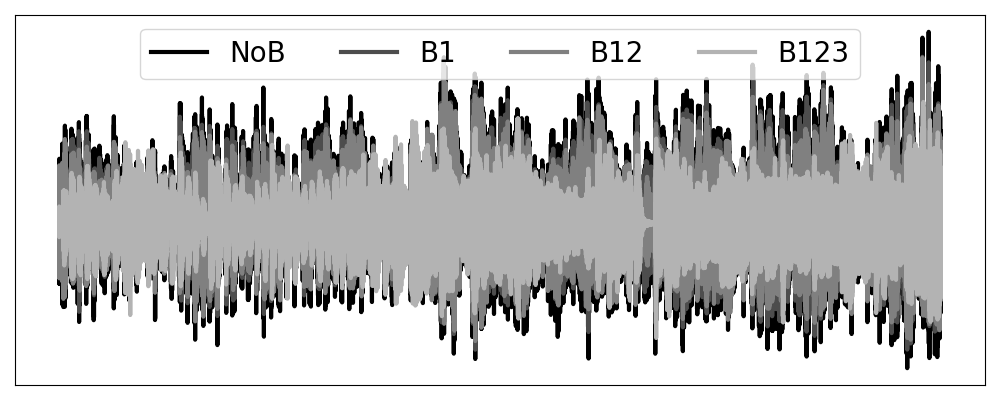}
		\caption{Other}
	\end{subfigure}%
	\caption{Waveform evolving as the blocks are returned (added) to a trained model. The model with no block, the one with block 1, the one with blocks 1 and 2, and the one with all blocks 1, 2 and 3 are represented in gray-scale shades from the darkest to the lightest shades. The `other' source and the other three soruces show opposite behaviors as the number of blocks increases.}
	\label{fig:block_evolving}
\end{figure*}

\subsection{How the Model Works}\label{sec:insight}

We investigate here how our model works. In a D2 block, the block input is added to the output of the last layer as shown in Figure \ref{fig:block_design}. This operation intuitively could fix the semantic of each channel, starting from the output of the first convolution to the output of the final block. To verify this intuition, we remove the blocks from the top of the stack one at a time in the trained dilation-2 model. This results in an altered model without any blocks, an altered model with block 1, an altered model with block 1 and 2, and the original model with block 1, 2 and 3. We apply these altered models to songs and convert them back to audios just like how we evaluate the full separation model described in Section \ref{sec:setting}. Note that these altered models use the same weights as the full model and are not re-trained.

We find that the output from these altered models have very good sound quality subjectively. In terms of separation performance, it gets better as more blocks are returned (added) to the model. The fact that the separations from these altered models are audible verifies the intuition that the semantics of the channels are fixed to some degree.

By plotting the waveforms converted from the outputs of the altered models, we can have a grasp of how the original model works. Figure \ref{fig:block_evolving} shows such plots for `other' and `vocals.' We observe that `vocals,' `drums,' and `other' get more and more activations as more blocks are used. In contrast, `other' gets less and less activations as more blocks are used. These observations give us some hints on the strategy the model has developed. In the lower blocks, the model stores most of the information in the `other' source. As the process moves forward, the `vocals' recovers more and more information from the channels related to the other three sources. The changes in `drums' and `bass' are relatively smaller compared to the change in `vocals.'  

The objective metrics also give us some hints on the process, as shown in Table \ref{tab:block_evolving}. `Drums' and `bass' are relatively simple signals compared to `vocals' and `other,' so the model can already roughly separate these two sources even without any blocks. In contrast, `vocals' and `other' are very poor without any blocks. The capacity of the model increases as the blocks are added back.

\begin{table}
\centering
\begin{tabular}{lrrrr}
\toprule
                & vocals& drums	& bass	& other\\
\midrule
No blocks       & 0.82  & 2.28  & 1.65  & -0.05 \\
Block 1         & 1.54  & 2.58  & 1.56  & 1.01  \\
Block 1, 2      & 2.82  & 3.69  & 2.86  & 1.85  \\
Block 1, 2, 3   & 6.74  & 5.71  & 5.00  & 4.61  \\
\bottomrule
\end{tabular}
\caption{Performance comparison (in SDR)  as the blocks are added to a trained model. In `No blocks,' all D2 blocks are removed from the model. Then, the blocks are added one by one.}
\label{tab:block_evolving}
\end{table}

\section{Conclusion}
We have presented a model for music source separation. It uses a stack of dilated convolutions as the backbone and consists of a stack of D2 blocks. In a D2 block, a Dilated GRU is combined with a dilated convolution to form a unit. Dilated GRU runs faster than standard GRU while still maintaining the performance. The proposed model achieves state-of-the-art performance in separating `vocals' and `accompaniment.'

Currently, music source separation focuses on separating sources in Pop music with vocals. In the future, we aim to separate different sources such as different instruments in symphony and rich electronic sounds in EDM.

%future edm, indivisual instruments

\bibliographystyle{named}
\bibliography{ijcai19}

\end{document}